\title{\bf Stability analysis of $ f(Q) $  gravity models using Dynamical Systems}
\author{Pooja Vishwakarma$^{1}$, Parth Shah$^{2}$ \\
$^{1}$Department of Mathematics, VIT AP University,\\
Amaravati 522237, India \\
$^{2}$Department of Mathematics, VIT AP University,\\
Amaravati 522237, India}

\documentclass[11pt,a4paper]{article}
\usepackage{amsmath, amsthm, amsfonts}
\usepackage[margin=1in,footskip=0.25in]{geometry}
\usepackage{makecell}
\usepackage[noadjust]{cite}
\usepackage{multicol}
\usepackage{graphicx}
\theoremstyle{theorem}

\usepackage{caption}
\usepackage{subcaption}
\newtheoremstyle{defi}
  {10pt}          
  {10pt}  
  {\rm}  
  {\parindent}     
  {\bf}  
  {. }    
  { }    
  {}     
\theoremstyle{defi}

\begin{document}

\date{}
    \maketitle
    \begin{abstract}
In recent years, the modified theory of gravity known as $f(Q)$ gravity has drawn interest as a potential alternative to general relativity. According to this theory, the gravitational force is determined by a function of the so-called ``non-metricity" tensor $Q$, which expresses how far a particle space-time is from the metric geometry. In contrast to general relativity, which describes the gravitational field using the curvature tensor, $f(Q)$ gravity builds a theory of gravity using the non-metricity tensor.
For this class of theories, dynamical system analysis of the background and perturbation equations has been carried out in this work to determine how various models behave cosmologically. Here, the critical points are determined for two $f(Q)$ models from the literature: the power law, $ f(Q)=Q+mQ^{n} $, and the logarithmic, $ f(Q)=\alpha+\beta log Q $ models. The stability behaviour and corresponding cosmology are displayed for each critical point. For the power law model, we achieve a matter-dominated saddle point with the right matter perturbation growth rate. For the logarithmic model, we get a saddle point dominated by the geometric component of the $ f(Q)$ model with perturbations in the decomposition of matter. For both models, we later achieved a stable and accelerating Universe with constant matter perturbations.
    \end{abstract}


\begin{multicols}{2}

\section{Introduction}
The Universe's present acceleration has been a hotly debated issue for the past 20 years. The Universe's expansion has begun to accelerate, according to recent observations \cite{1, 2, 3, 4, 5, 6, 7} of type Ia supernovae (SNIa), Large-Scale Structure (LSS), Wilkinson Microwave Anisotropy Probe (WMAP) data, Cosmic Microwave Background (CMB), and Baryonic Acoustic Oscillations (BAO).  The same observational data also shows that everything we currently see makes up only $4$ percent of the Universe's total mass, with the other $96$ percent consisting of mysterious substances named Dark Energy (DE), which makes up about $70$ percent and Dark Matter (DM), which makes up about $26$ percent. The outcomes of these recent observations go counter to the standard Friedmann equations of general relativity (GR), which are part of GR's large-scale applications to a homogeneous and isotropic Universe. GR may be a specific instance of a more general theory rather than the final theory of gravitation, as we would have imagined \cite{8}. The major problem surfaces when one is trying to address the Universe's acceleration theoretically. 

Two theories in the literature aim to explain why the Universe is expanding so rapidly right now \cite{9}. The first strategy in the GR framework is to alter the composition of the Universe by introducing new elements of matter and energy, such as DE of an as-of-yet unidentified type that has a significant negative pressure. The popular candidate for DE is the well-known cosmological constant $\Lambda$, which Einstein included in his field equations in GR and which closely matches the previously described evidence. This is the incredibly straightforward theory known as the $\Lambda CDM$ model, which can account for the Universe's cosmic acceleration in the late cosmos. Modifying the Einstein-Hilbert action in GR theory is the second strategy for explaining the current acceleration of the Universe's expansion. The so-called $f(R)$ gravity \cite{10,11,12}, which generalizes the Hilbert-Einstein action to a generic function of the Ricci Scalar $R$, is the simplest form of modification of GR. The Ricci scalar $R$, which describes the curvature imported from Riemannian geometry, is the essential idea in general relativity (GR). This modified theory of gravity is renowned for accurately representing cosmic acceleration. It can also accurately recreate the entire history of the cosmos and the behaviour of the cosmological constant ($\Lambda $). Other modified theories include  the $f(R,T)$ theory \cite{13,14}, $f(R,L_{m})$ theory \cite{15,16}, where $L_{m}$ is the Lagrangian density, and the $f(R,G)$ theory \cite{17,18}, where $G$ denotes the Gauss-Bonnet invariant, among many more.

There are further alternatives to general relativity (GR), such as the teleparallel equivalent of GR (TEGR), which uses the concept of torsion $T$ to describe gravitational interactions. Although the Weitzenbock link in teleparallelism is related to torsion with zero curvature\cite{19}, the Levi-Civita connection in general relativity is associated with curvature with zero torsion. The $f(T)$ gravity is also the most straightforward adaptation of TEGR \cite{20,21}. The symmetric teleparallel equivalent of general relativity (STEGR), a novel theory of gravity, has recently been put out. In this theory, the gravitational interactions are defined by the idea of non-metricity $Q$ with zero torsion and curvature \cite{22,23}. Weyl geometry (a generalization of Riemannian geometry non-metricity) characterizes the variation in a vector's length in parallel transit and serves as the main geometrical factor explaining the characteristics of gravitational interaction. In Weyl geometry, $Q_{\gamma \mu \nu}=-\bigtriangledown_{\gamma}g_{\mu \nu}$ \cite{24}, is used to mathematically derive the covariant derivative of the metric tensor, which is not equal to zero. The $f(Q)$ gravity is additionally STEGR's most straightforward adaptation. Torsion and non-metricity, respectively, provide fascinating geometric frameworks on which teleparallel and symmetric teleparallel gravity can be constructed \cite{25,26,27,28,29,30}. 

Due to non-linear components in the field equations, one of the main issues with theories of gravity is the difficulty in obtaining solutions (analytical or numerical), making it impossible to compare them with observations. The method known as ``Dynamical System Analysis" is used to resolve such equations or regulate overall dynamical behaviour. This strategy seeks out numerical answers to better comprehend the qualitative behaviour of a particular physical system \cite{10,31,32,33,34,35,36,37,38,39}. Finding the critical points of a collection of first-order ordinary differential equations is the most crucial idea in dynamical system analysis. By computing the Jacobian matrix at critical points and determining its eigenvalues, the stability criteria are determined. After locating the critical points and eigenvalues, we can linearize the system around each point to determine the flow around that point. Studying stability characteristics close to a specific critical point is what this is about \cite{40,41,42,43,44}.

In this work, we examine the cosmic dynamics of $f(Q)$ gravity models at both background and perturbation levels using the potent mathematical tool of dynamical system analysis, which is motivated by the fascinating properties of $f(Q)$ gravity. Such research can be used to strengthen the conclusions drawn from the observational analysis. We point out that the dynamical system technique is typically used at the background level, but it can also be used at the perturbation level \cite{45,46,47,48,49,50,51,52,53,54,55,56,57,58,59,60}. This realization was made recently \cite{61,62,63,64}. So, regardless of the precise initial conditions, we may determine the background stable late-time solutions as well as the growth of the structure creation using this combined approach. Additionally, we can look at the manner in which the matter perturbations respond to the background solutions. The fact that $f(Q)$ gravity automatically incorporates a preferred frame into the theory is one of the main reasons to take it into consideration. The work is structured as follows: We state the $f(Q)$ gravity field equations in Section II, from which the background and perturbed cosmological equations can be derived. The dynamical study of the combined system for the power-law and exponential models is covered in Section III. In Section IV, the final results are summarised.

\section{$f(Q)$ COSMOLOGY}
Jimenez et al. \cite{22,23,65,66} introduced the symmetric teleparallel gravity, often known as the $f(Q)$ gravity. This modified gravity's action is provided by
\begin{equation}
S=\int [-\frac{1}{16\pi G}f(Q)+{\mathcal{L}}_{m}]\sqrt{-g} \,d^{4}x\ ,
\end{equation}
where $g$ is the determinant of the metric tensor $ g_{\mu \nu} $, and $ {\mathcal{L}}_{m} $ is Lagrangian density of matter. $ f(Q) $ is an arbitrary function of $ Q $, the non-metricity scalar which is responsible for the gravitational interaction. 

The non-metricity tensor is given by
\begin{center}
$ Q_{\lambda \mu \nu}=\nabla_{\lambda}g_{\mu \nu}=\partial_{\lambda}g_{\mu \nu}-\Gamma^{\beta}_{\lambda \mu}g_{\beta \nu}-\Gamma^{\beta}_{\lambda \nu}g_{\mu \beta} ,$ 
\end{center}
where $ \Gamma^{\beta}_{\mu \nu} $ is the affine connection and it can be decomposed into following components:
\begin{center}
$\Gamma^{\beta}_{\mu \nu}=\lbrace^{\beta}_{\mu \nu}\rbrace+K^{\beta}_{\mu \nu}+L^{\beta}_{\mu \nu}$ ,
\end{center}
where the Levi-Civita connection $ \lbrace^{\beta}_{\mu \nu}\rbrace $ is defined by 
\begin{center}
$ \lbrace^{\beta}_{\mu \nu}\rbrace=\frac{1}{2}g^{\beta \sigma}(\partial_{\mu}g_{\sigma \nu}+\partial_{\nu}g_{\sigma \mu}-\partial_{\sigma}g_{\mu \nu})$ ,
\end{center}
$ L^{\beta}_{\mu \nu} $ is the deformation defined by
\begin{center}
$L^{\beta}_{\mu \nu}=\frac{1}{2}Q^{\beta}_{\mu\nu}-Q_{(\mu^{\beta}\nu)}$ ,
\end{center}
 and $ K^{\beta}_{\mu \nu} $ is the contortion defined by
\begin{center}
$K^{\beta}_{\mu \nu}=\frac{1}{2}T^{\beta}_{\mu\nu}+T_{(\mu^{\beta}\nu)}$ ,
\end{center}
with the torsion tensor $T^{\beta}_{\mu\nu} $ defined as the anti-symmetric part of the affine connection, 
\begin{center}
$T^{\beta}_{\mu \nu}=2\Gamma^{\lambda}_{[\mu \nu]} $ .
\end{center}
The non-metricity scalar $Q$ is written as \cite{66,67} 
\begin{equation}
\begin{split}
& Q=-\frac{1}{4}Q_{\alpha \beta \gamma}Q^{\alpha \beta \gamma}+\frac{1}{2}Q_{\alpha \beta \gamma}Q^{\gamma \beta \alpha}\\
& + \frac{1}{4}Q_{\alpha}Q^{\alpha}-\frac{1}{2}Q_{\alpha}\bar{Q}^{\alpha} ,
\end{split}
\end{equation} where \begin{center}
$ Q_{\alpha}\equiv Q_{\alpha^{\mu}\mu} $ and $ \bar{Q}^{\alpha}\equiv Q_{\mu}^{\mu \alpha} $
\end{center} 
consist of two independence traces that were obtained by condensing the non-metricity tensor
\begin{center} $ Q_{\alpha \mu \nu}\equiv \bigtriangledown_{\alpha}g_{\mu \nu} $ , \end{center} 
which enables us to define the non-metricity scalar as $ Q=-Q_{\alpha \mu \nu } Q $.

To find the field equations, we set that action in (1) is constant with respect to variations over the metric tensor $ g_{\mu \nu} $ and setting $ 8\pi G=1 $ for simplicity, results in \cite{67,68} 
\begin{equation}
\begin{split}
& \frac{2}{\sqrt{-g}}\bigtriangledown_{\alpha}\sqrt{-g}g_{\beta \nu}f_{Q}[-\frac{1}{2}L^{\alpha \mu \beta}+\frac{1}{4}g^{\mu \beta}(Q^{\alpha}-\bar{Q}^{\alpha})\\
& -\frac{1}{8}(g^{\alpha \mu}Q^{\beta}+g^{\alpha \beta}Q^{\mu})]+f_{Q}[-\frac{1}{2}L^{\mu \alpha \beta}-\frac{1}{8}(g^{\mu \alpha}Q^{\beta} \\
& +g^{\mu \beta}Q^{\alpha})+ \frac{1}{4}g^{\alpha \beta}(Q^{\alpha}-\bar{Q}^{\alpha})]Q_{\nu \alpha \beta}+\frac{1}{2}\delta_{\nu}^{\mu}f =T^{\mu}_{\nu} ,
\end{split} 
\end{equation}
where $ T_{\mu \nu} $ is the matter energy-momentum tensor , whose form is \begin{center}
$ T_{\mu \nu}=-\frac{2}{\sqrt{-g}}\frac{\delta(\sqrt{-g}{\mathcal{L}}_{m})}{\delta g^{\mu \nu}} $ ,
\end{center}
and $ f_{Q}=\frac{\partial f}{\partial Q} $.

At the background level, we assume a Friedmann-Lemaitre-Robertson-Walker (FLRW) spacetime that is homogeneous, isotropic, and spatially flat, whose metric is of the form \begin{equation}
ds^{2}=-N^{2}(t)dt^{2}+a^{2}(t)(dx^{2}+dy^{2}+dz^{2}) ,
\end{equation}
where $ N(t)$ is a Lapse function and $ t $ is the cosmic time. Because $Q$ theories retain a residual time reparameterization invariant, they allow for the selection of a specific Lapse function. As a result, symmetry is used to set $N(t)=1$ \cite{22,23,69,70}. The function $ a(t) $ is known as the scale factor, is a measure of the size of the Universe at time $t$ and $ x , y , z $ are the cartesian coordinates. Note that for the non-metricity scalar, we obtain $ Q=6H^{2} $ in FLRW metric, where $ H=\frac{\dot{a}}{a} $ is the Hubble function which is a measure of the Universe's expansion rate at time $ t $ and `.' denotes derivative with respect to $ t $. Imposing the splitting $ f(Q)=Q+F(Q) $, and applying the FLRW metric, the corresponding field equations are \cite{66,67,68,69,70,71}
\begin{equation}
3H^{2}=\rho + \frac{F}{2}-QF_{Q} ,
\end{equation}
\begin{equation}
(2QF_{QQ}+F_{Q}+1)\dot{H}+\frac{1}{4}(Q+2QF_{Q}-F)=-2p ,
\end{equation}
where $ F_{Q}=\frac{dF}{dQ} $ , $ F_{QQ}=\frac{d^{2}F}{dQ^{2}} $ , $ \rho $ and $ p $ are the matter fluid's energy density and pressure respectively. They satisfy the energy conservation equation or the fluid equation in the absence of interaction which is given by
\begin{equation}
\dot{\rho}+3H(\rho+p)=0
\end{equation}
with a linear equation of state parameter between them given by $ p=\rho \omega $, where
the equation of state parameter  $ \omega\in [-1,1] $. From the Friedmann's equation (5), we have
\begin{center}
$ 1=\frac{\rho}{3H^{2}} +\frac{\frac{F}{2}-QF_{Q}}{3H^{2}} $.
\end{center}
For an accelerating Universe, one requires $ \omega_{eff}<-\frac{1}{3} $ \cite{10,71} and therefore, it is convenient to introduce the energy density parameters $ \Omega_{m} $ and $ \Omega_{Q} $ as
\begin{equation}
\Omega_{m}=\frac{\rho}{3H^{2}} ,\  \Omega_{Q}=\frac{\frac{F}{2}-QF_{Q}}{3H^{2}} .
\end{equation}
Hence the Friedmann's equation(5) can be simply written as
\begin{equation}
\Omega_{m}+\Omega_{Q}=1 .
\end{equation}

Introducing the effective total energy density $ \rho_{eff} $ and total energy pressure $ p_{eff} $, as \cite{70}
\begin{equation}
\rho_{eff}\equiv \rho+\frac{F}{2}-QF_{Q}  ,
\end{equation}
\begin{equation}
p_{eff}\equiv \frac{\rho (1+\omega)}{2QF_{QQ}+F_{Q}+1}-\frac{Q}{2}  ,
\end{equation}
the corresponding total equation of state $ \omega_{eff} $ can be written as
\begin{equation}
\omega_{eff}=\frac{p_{eff}}{\rho_{eff}}=-1+\frac{\Omega_{m}(1+\omega)}{2QF_{QQ}+F_{Q}+1} .
\end{equation}
For the investigation of the linear perturbation level, we focus on the matter density contrast $ \delta=\frac{\delta _{\rho}}{\rho} $, where $ \delta_{ \rho} $ is the perturbation of the matter energy density. The evolution equation for matter over density at the quasi-static limit \cite{67,68,69} is
\begin{equation}
\ddot{\delta}+2H\dot{\delta}=\frac{\rho \delta}{2(1+F_{Q})} ,
\end{equation}
where the right hand side's denominator explains how an apparent Newton's constant appears. The terms involving temporal derivatives in the perturbed equations are omitted in the quasi-static limit at tiny scales, well within the cosmic horizon, leaving just spatial derivative terms \cite{10,71,72}.

\section{DYNAMICAL SYSTEM ANALYSIS OF THE MODELS}
For a generic function $ F(Q) $, we build the dynamical system of the background and perturbed equations in this section. In this regard, we transform the equations (5)-(7) and (13) into first order autonomous system by taking into account the following dynamical variables:
\begin{equation}
 x=\frac{F}{6 H^{2}}, \  y=-2 F_{Q},\  u=\frac{d(\ln \delta)}{d(\ln a)} .
\end{equation}
As a result, while variables $x$ and $y$ are related to Universe's background evolution, variable $ u $ measures the expansion of matter disturbances. Hence, at any point of time the matter density contrast is positive. $ u>0 $ denotes the increase of matter perturbations and $ u<0 $ denotes their decay.

The cosmic background parameters $\Omega_{m}$ , $\Omega_{Q}$, and $\omega_{\text {eff }}$ can be represented as \cite{65}
\begin{center}
$ \Omega_{m}=1-x-y $ ,
\end{center}
\begin{center}
$ \Omega_{Q}=x+y $ ,
\end{center}
\begin{equation}
\omega_{eff }=-1+\frac{(1-x-y)(1+\omega)}{2QF_{QQ}-\frac{y}{2}+1} .
\end{equation}
The cosmological equations can now be expressed as the following dynamical system in terms of variables (14):
\begin{equation}
x'=-\frac{\dot{H}}{H^{2}}(y+2x) ,
\end{equation}
\begin{equation}
y'=2y\frac{\dot{H}}{H^{2}}\frac{QF_{QQ}}{F_{Q}} ,
\end{equation}
\begin{equation}
u'=-u(u+2)+\frac{3(1-x-y)}{2-y}-\frac{\dot{H}}{H^{2}}u ,
\end{equation}
where $ ( ' ) $ denotes the differentiation with respect to $ ln a $ and $ ( . ) $ stands for differentiation with respect to $ t $ and
\begin{equation}
\frac{\dot{H}}{H^{2}}=-\frac{3-3(1-x-y)}{4QF_{QQ}-y+2} .
\end{equation}
The physical system is a product space of the perturbed space $ \Bbb P $, which contains the variable $ u $ , and the background phase space $ \Bbb B $, which contains the variable $ x $ and $ y $. The combined system's phase space under the physical condition $ 0\leq \Omega_{m}\leq 1 $ is
\begin{center}
$ \Psi= \Bbb B \times \Bbb P = (x,y,u)\in \Bbb R^{2}\times \Bbb R : 0\leq x+y \leq 1 .$ 
\end{center}
Note that the projection of orbits of the product space $ \Psi $ on space $ \Bbb B $ reduces to the corresponding background orbits.

The system's dynamical evolution will be described in the following stage by isolating its critical points and evaluating their stability. In terms of physics, a stable point with $ u>0 $ denotes an unbounded expansion of matter perturbations, and as a result, the system is not stable with respect to matter perturbations. The system is asymptotically stable with respect to perturbations because a stable point with $ u<0 $ denotes the decay of mater disturbances. Finally, it is implied that matter perturbations remain constant when $ u=0 $ at a stable position. In conclusion, the perturbations of matter expand but do not last forever in an unstable or saddle point with $u>0$, which is required to explain the matter epoch of the universe. This unstable or saddle point must be followed by a stable late-time attractor with $u=0$, which corresponds to acceleration \cite{65}.

We must define the function $ F $ in order to move on to the specific analysis and, as a result, establish the term $ \frac{QF_{QQ}}{F_{Q}} $. In the subsections that follow, we discuss two specific models that are known to provide an intriguing cosmic phenomenology.
\subsection*{MODEL I : $ F(Q)=mQ^{n} $}
We begin by taking into account a power-law model with
\begin{equation}
 F(Q)=mQ^{n} 
\end{equation}
having two parameters, $ m $ and $ n $. This model fits the BBN requirement \cite{73} and can explain the acceleration of the Universe in the late Universe. We point out that this model reduces to the symmetric teleparallel equivalent of general relativity for $ n=1 $ and that for $ n=0 $ it is equivalent to the concordance $\Lambda CDM$ scenario \cite{67,74,75}. As
\begin{center}
$ QF_{QQ}=\frac{(1-n)}{2}y $
\end{center}
in this instance, the system (16)-(18) becomes,
\begin{equation}
x'=\frac{3(1-x-y)}{y(1-2n)+2}(y+2x) ,
\end{equation}
\begin{equation}
y'=\frac{3(1-x-y)}{y(1-2n)+2}2(1-n)y ,
\end{equation}
\begin{equation}
u'=\frac{3(1-x-y)}{2-y}-u(u+2)+\frac{3(3-x-y)}{y(1-2n)+2}u .
\end{equation}

Four critical points can be found in the relevant dynamical system:
\end{multicols}
\begin{table}[h!]
\centering
\begin{tabular}{ |p{3cm}||p{3.5cm}||p{1cm}||p{1cm}||p{1cm}||p{4cm}|| }
 \hline
 \textbf{Critical Points} & \textbf{Eigen-values} & $\mathbf{\Omega_{m}}$ & $\mathbf{\omega_{eff} }$ & $\mathbf{u}$ & \textbf{Stability condition}\\
 \hline
  $ (1-y,y,0) $ & $ (0,-2,-3) $ & $ 0 $ & $ -1 $ & $ 0 $ & Stable\\
 \hline
 $ (1-y,y,-2) $ & $ (0,2,-3) $ & $ 0 $ & $ -1 $ & $ -2 $ & Saddle\\
 \hline
$ (0,0,-\frac{3}{2}) $ & $ (3,\frac{5}{2},-3(1-n)) $ & $ 1 $ & $ 0 $ & $ -\frac{3}{2} $ & Unstable for $ n<1 $, Saddle for $ n>1 $\\
 \hline
$ (0,0,1) $ & $ (3,-\frac{5}{2},-3(1-n)) $ & $ 1 $ & $ 0 $ & $ 1 $  & Saddle\\
 \hline
\end{tabular}
\caption{Critical points, Stability conditions, EoS parameter and matter perturbation}
\label{1}
\end{table} 

\begin{figure}[!htp]   
\centering    
    \mbox{\includegraphics[scale=0.53]{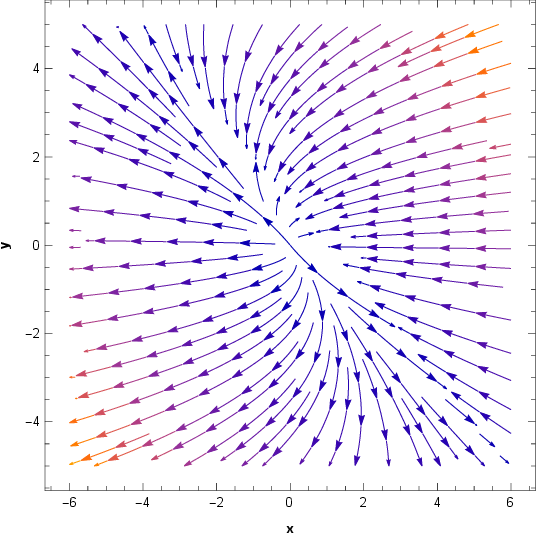}}   
    \hspace{10px}
    \mbox{\includegraphics[scale=0.53]{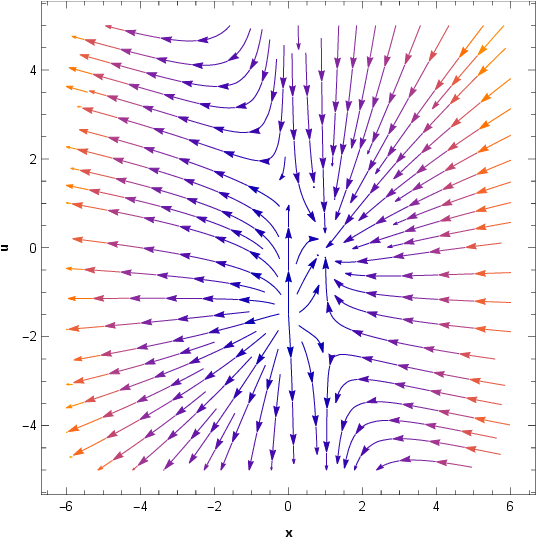}}
    \hspace{10px}
    \mbox{\includegraphics[scale=0.53]{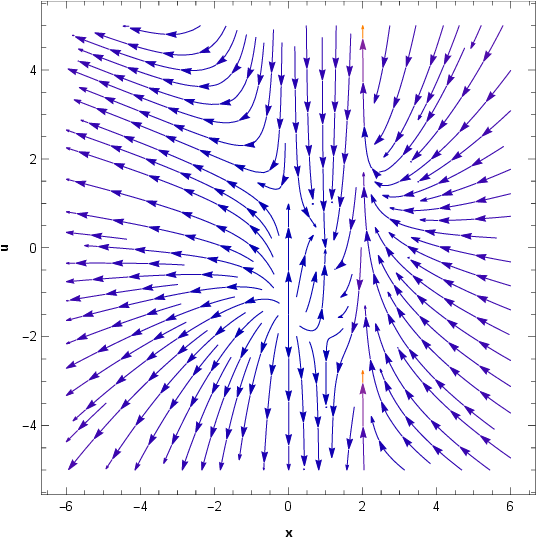}}
    \caption{ Phase Portrait for the dynamical system of Model I, (i)left panel($u=0$); (ii)middle panel($y=0$); (iii)right panel($x=0$ and $n=0.5$)}
    \label{Phase Portrait for the dynamical system of Model I}
\end{figure}

\begin{multicols}{2}
\textbf{* Critical Point $(1-y,y,0)$ :} This point's curve represents a solution in which the effective dark energy component dominates, or $ \Omega_{Q}=1 $, and the Universe accelerates with $ \omega_{eff}=-1 $, which is similar to the behaviour of a cosmological constant. In this case, it indicates that the critical point is dominated by the geometric component of $ f(Q) $ model. Moreover, $ u=0 $ at the perturbation level means that the matter perturbation is still constant. The curve is one-dimensional with one vanishing eigenvalue, and other eigenvalues are $ -2 $ and $ -3 $. One may assess the stability of the curve by looking at the signature of the remaining non-vanishing eigenvalues. Hence, we draw the conclusion that it is consistently stable. In addition, there is one vanishing eigenvalue in the corresponding Jacobian matrix, which is the non-hyperbolic point. As a result, we discover that the system of equations is asymptotically stable at its origin using the Center Manifold Theory.

In this case $x$ is the central variable and $y,u$ are the stable variables. The corresponding matrices are
$ A=0 $ and $ B= \begin{pmatrix}
-2 & 0\\
0 & -3
\end{pmatrix} $. The center manifold has now the form $ y=h_{1}(x) $ and $ u=h_{2}(x) $; the approximation $ N $ has two components:
\begin{flushleft}
$ N_{1}(h_{1}(x))=h_{1}'(x)\frac{3(1-x-y)}{y(1-2n)+2}(y+2x) -\frac{3(1-x-y)}{y(1-2n)+2}2(1-n)y ,$
\end{flushleft}
\begin{flushleft}
$ N_{2}(h_{2}(x))=h_{2}'(x)\frac{3(1-x-y)}{y(1-2n)+2}(y+2x)-\frac{3(1-x-y)}{2-y}+u(u+2)-\frac{3(1-x-y)}{y(1-2n)+2}u .$
\end{flushleft}
For zeroth approximation:\\ $ N_{1}(h_{1}(x))=0+\mathcal{O}(x^{2}) $ and\\ $ N_{2}(h_{2}(x))=-\frac{1}{2}(1-x)+\mathcal{O}(x^{2}) $.
\\Therefore the reduced equation gives us
\begin{flushleft}
$ u'=-6+6x+O(x^{2}) $
\end{flushleft}
This gives us the negative constant part.
Therefore, the system of equations (21)-(23) is asymptotically stable at zero as per the Central Manifold Theory.

In conclusion, the argument outlines the dark-energy-dominated Universe of the late Universe at background and perturbation levels.
 
\textbf{* Critical Point $ (1-y,y,-2) $: }The curve of this critical point $ (1-y,y,-2) $, like the curve of critical point $ (1-y,y,0) $, corresponds to a solution with $ \omega_{eff}=-1 $, which is dominated by the geometric component of $ f(Q) $ model. Also, $ u=-2 $ implies that it is characterised by the decay of matter disturbances. However, it is  saddle with eigenvalues $ (0,2,-3) $. As a result, unlike the curve of the critical point $ (1-y,y,0) $, the critical point $ (1-y,y,-2) $ cannot, at the perturbation level, reflect a late-time, dark-energy-dominated Universe. This point depicts the inflationary era of the Universe due to its saddle nature and negative $\omega_{eff}$ value. 

\textbf{* Critical Point $ (0,0,-\frac{3}{2}) $: }This position, where $ \Omega_{m}=1 $ and $ \omega_{eff}=0 $, corresponds to the matter supremacy at the background level. This statement however, could not explain the development of structures at the perturbation level because the matter over density fluctuates as $\rho\propto a^{-\frac{3}{2}} $ when $ u=-\frac{3}{2} $. The Jacobian matrix's eigenvalues are  $ (3,\frac{5}{2},-3(1-n)) $, making this point unstable for $ n\geq 1 $ and saddle for $ n<1 $.

\textbf{* Critical Point $ (0,0,1) $: }The background parameters are $ \Omega_{m}=1 $ and $ \omega_{eff}=0 $, and this point corresponds to a matter-dominated critical solution. We have $ u=1 $ at the perturbation level, which means that the matter over density $ \delta $ varies as the scale factor $ \rho\propto a $ and hence rises with the expansion of the Universe. Given that the appropriate Jacobian matrix has the eigenvalues $ (3,-\frac{5}{2},-3(1-n)) $, the point $ (0,0,1) $ is always a saddle point for any value of $ n $. In light of this, the trajectories pass through this point and then depart from it after being drawn to a late-time stable point. As a result, we come to the conclusion that this point may be the best choice to explain how structures were formed during the matter-dominant period at both the background and perturbation levels. 

\subsection*{MODEL II : $ F(Q)=\alpha+\beta log Q-Q $}
We consider the logarithmic model in this subsection
\begin{equation}
F(Q)=\alpha+\beta logQ-Q ,
\end{equation}
which has two parameters, $ \alpha $ and $ \beta $. This model also fits in the BBN requirements \cite{73} and can explain the acceleration of the Universe in the late Universe. As, 
\begin{center}
$  QF_{QQ}=\frac{(y-2)}{2}$
\end{center}
in this instance, the system (16)-(18) becomes.
\begin{equation}
x'=\frac{3(1-x-y)}{y-2}(y+2x) ,
\end{equation}
\begin{equation}
y'=\frac{3(1-x-y)}{y-2}2(y-2) ,
\end{equation}
\begin{equation}
u'=\frac{3(1-x-y)}{2-y}-u(u+2)+\frac{3(1-x-y)}{y-2}u .
\end{equation}

Two critical points can be found in this dynamical system:
\end{multicols}

\begin{table}[h!]
\centering
\begin{tabular}{ |p{3cm}||p{3.5cm}||p{1cm}||p{1cm}||p{1cm}||p{4cm}|| }
 \hline
    \textbf{Critical Points} & \textbf{Eigen-values} & $\mathbf{\Omega_{m}} $ & $\mathbf{ \omega_{eff}} $ & $ \mathbf{u} $ & \textbf{Stability condition}\\
 \hline
  $ (1-y,y,0) $ & $ (0,-3,-2) $ & $ 0 $ & $ -1 $ & $ 0 $ & Stable\\
 \hline
 $ (1-y,y,-2) $ & $ (0,-3,2) $ & $ 0 $ & $ -1 $ & $ -2 $ & Saddle\\
 \hline
\end{tabular}
\caption{Critical points, Stability conditions, EoS parameter and matter perturbation}
\label{2}
\end{table} 

\begin{figure}[!htp]   
\centering    
    \mbox{\includegraphics[scale=0.53]{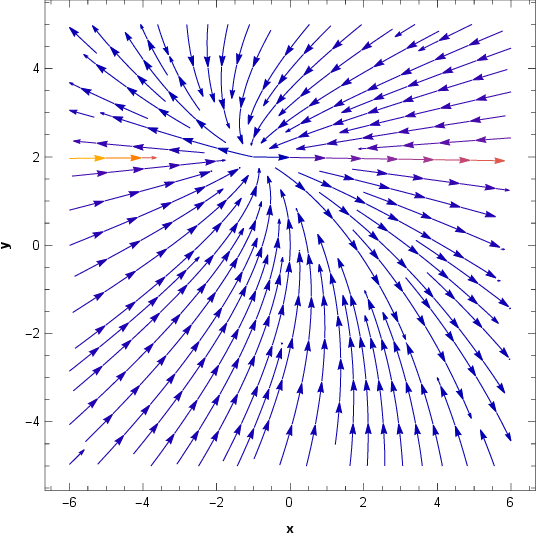}}   
    \hspace{10px}
    \mbox{\includegraphics[scale=0.53]{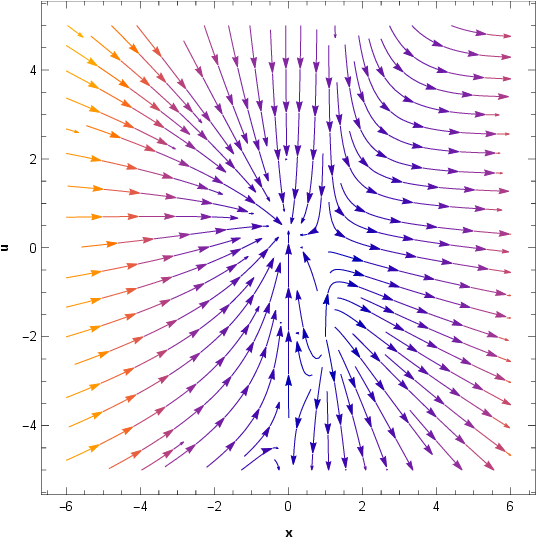}}
    \hspace{10px}
    \mbox{\includegraphics[scale=0.53]{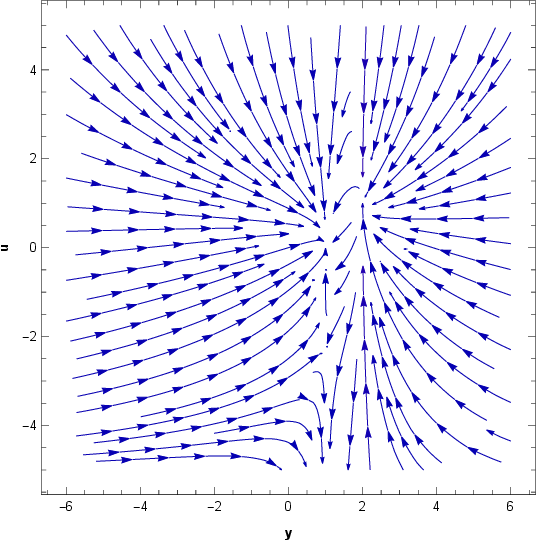}}
    \caption{ Phase Portrait for the dynamical system of Model II, (i)left panel($u=0$); (ii)middle panel($ y=0 $); (iii)right panel($ x=0 $)}
    \label{Phase Portrait for the dynamical system of Model II}
\end{figure}

\begin{multicols}{2}
\textbf{* Critical Point $ (1-y,y,0) $: } The curves at this time depict a solution in which the Universe accelerates with $ \omega_{eff}=-1 $ and $ \omega_{Q}=1 $, which is consistent with the behaviour of a cosmological constant dominant Universe. Moreover, the matter perturbation is still constant because $ u=0 $ at the perturbation level. The corresponding eigenvalues of the one-dimensional curve are $ (0,-2,-3) $. By examining the signature of the non-vanishing eigenvalues one can judge the stability of the curve. So, we infer that it is consistently stable. Since it is a non-hyperbolic point due to one vanishing eigenvalue, we use the Center Manifold Theory to study the system of equations.

Here $x$ is the central variable and $ y, u$ are the stable variables. The corresponding matrices are
$ A=0 $ and $ B= \begin{pmatrix}
-3 & 0\\
0 & -2
\end{pmatrix} $. The center manifold has now the form $ y=h_{1}(x) $ and $ u=h_{2}(x) $; the approximation $ N $ has two components:
\begin{flushleft}
$ N_{1}(h_{1}(x))=h_{1}'(x)\frac{3(1-x-y)}{y-2}(y+2x) -\frac{3(1-x-y)}{y-2}2(y-2) , $
\end{flushleft}
\begin{flushleft}
$ N_{2}(h_{2}(x))=h_{2}'(x)\frac{3(1-x-y)}{y-2}(y+2x)-\frac{3(1-x-y)}{2-y}+u(u+2)-\frac{3(1-x-y)}{y-2}u .$
\end{flushleft}
For zeroth approximation:\\ $ N_{1}(h_{1}(x))=6(x-1)+\mathcal{O}(x^{2}) $ and\\ $ N_{2}(h_{2}(x))=-\frac{3}{2}(1-x)+\mathcal{O}(x^{2}) $.
\\Therefore the reduced equation gives us
\begin{flushleft}
$ u'=-\frac{9}{4} \frac{(x-1)}{(3x-4)}(3x^{2}-10x+1)+\mathcal{O}(x^{2}) .$
\end{flushleft}
This gives us the negative constant part.
Therefore, the equation (25)-(27) is asymptotically stable at zero.

This point describes late Universe's dark-energy-dominated background and perturbation levels.

\textbf{* Critical Point $ (1-y,y,-2) $: } The curve of this critical point $ (1-y,y,-2) $ is similar to the curve of the critical point $ (1-y,y,0) $, and both curves corresponds to a solution with $ \omega_{eff}=-1 $, which is dominated by the geometric component of $ f(Q) $ model. In addition, it exhibits perturbations in the decomposition of matter because $ u=-2 $. Eigenvalues $(0,2,-3)$ are present, though. Thus, this critical point $ (1-y,y,-2) $ cannot, at the perturbation level, depict a late-time dark-energy-dominated Universe, in contrast to the curve of critical points $ (1-y,y,0) $. This point depicts the inflationary era of the Universe due to its saddle nature and negative $\omega_{eff}$ value.

\section{Discussion and Conclusion}
In the current work, we performed a combined dynamical system analysis of both background and perturbation equations in order to examine the validity of each result using an independent method. The Universe is assumed to be described by the flat FLRW metric. This work was motivated by the fact that cosmological models based on $f(Q)$ gravity are very efficient in fitting observational datasets at both background and perturbation levels.

After converting the background and perturbation equations into an autonomous system, we concentrated on the power-law and exponential $f(Q)$ models that have been examined in the literature. Due to the additional variable associated with matter perturbations, each background critical point splits into two points, each with a distinct behaviour of matter perturbations and stability.

In the power-law model $ f(Q)=Q+mQ^{n} $, there were four critical points, Out of which $(1-y,y,0)$ point was found to be asymptotically stable at origin by using Center Manifold Theory. The points $ (1-y,y,-2) $ and $ (0,0,1) $ were saddle points where matter predominated and the growth rate of matter perturbations was accurate. This was followed by a change to a stable, accelerated Universe where dark energy predominated and the growth rate of matter perturbations was constant, as shown through the critical points in Table 1. The point $ (0,0,-\frac{3}{2}) $ was unstable for $ n\geq 1 $ and saddle for $ n<1 $. We also plotted phase portraits in Figure 1, where the behaviour at those points is more clearly represented. Our study shows that various critical points correspond to various kinds of matter disturbances. Furthermore, similar background critical points react in various ways at the perturbation level. The points $ (0,0,1) $ and $ (0,0,-\frac{3}{2}) $ describe the background level of the decelerated matter-dominated era, but only point $ (0,0,1) $ exhibits the proper growth of matter structure. Also, Point $ (0,0,1) $ is a saddle, therefore it provides a natural exit to a late-time accelerated period. However, at late times, defining the accelerated dark-energy dominant epoch, the curves of critical points $(1-y,y,0)$ and $ (1-y,y,-2) $ are identical at the background level. However, only curve $(1-y,y,0)$ displays continuous matter perturbations and is stable, making it the only curve at the perturbation level that is physically and observably interesting. The model describes the transition from matter domination towards an accelerated dark-energy-dominated epoch. In conclusion, the power-law model is capable of capturing both the background and perturbation levels of the required thermal history of the universe. According to our investigation, the aforementioned is true for any value of n.

In the logarithmic model $ f(Q)=\alpha+\beta log Q $, there were two critical points $(1-y,y,0)$ and $ (1-y,y,-2) $. Using the Center Manifold Theory, the point $(1-y,y,0)$ was found to be asymptotically stable at the origin. We also obtained curves of points that correspond to matter dominance, the geometric component of $ f(Q) $ model dominance, and matter perturbation growth. The fact that these curves are saddle-shaped allows a smooth transition to the stable solution with constant matter perturbations, as demonstrated by the critical points in Table 2. In Figure 2, we also showed phase portraits that more accurately depict the behaviour at those points. Similar to Model I, we observe that the distinction between critical points that are similar at the background level is made by the presence of disturbances. Because of this, we discover through the combined background and perturbation analysis that only curve $(1-y,y,0)$ corresponds to late-time dark-energy dominance, with a stable history of matter perturbations. 
As a final remark, in this article, stability and acceleration phases are achieved without adding an ``unobserved"  quantity, i.e., dark energy.

\end{multicols}


\begin{thebibliography}{}
\bibitem{1}
Perlmutter S.; Aldering G.; Goldhaber G.; Knop R.A.; Nugent P.; Castro P.G.; Deustua S.; Fabbro S.; Goobar A.; Groom D.E.; et al. [The Supernova Cosmology Project]. Astrophys. J. 1999, 517, 565-586.
\bibitem{2}
Riess A.G.; Filippenko A.V.; Challis P.; Clocchiatti A.; Diercks A.; Garnavich P.M.; Gilliland R.L.; Hogan J.C.; Jha S.; Kirshner P.R.; et al. Astrophys. J. 1998, 116, 1009–1038.
\bibitem{3}
Riess A.G.; Strolger L.G.; Tonry J.; Casertano S.; Ferguson H.C.; Mobasher B.; Challis P.; Filippenko A.V.; Jha S.; Li W.; et al. Astrophys. J. 2004, 607, 665–687.
\bibitem{4}
Spergel D.N.; Verde L.; Peiris H.V.; Komatsu E.; Nolta M.R.; Bennett C.L.; Halpern M.; Hinshaw G.; Jarosik N.; Kogut A.; et al. Astrophys. J. Suppl. 2003, 148, 175–194.
\bibitem{5}
Koivisto T.; Mota D.F. Phys. Rev. D 2006, 73, 083502.
\bibitem{6}
Daniel S.F.; Caldwell R.R.; Cooray A.; Melchiorri A. Phys. Rev. D 2008, 77, 103513.
\bibitem{7}
Gadbail G.N.; Mandal S.; Sahoo P.K. Physics 2022, 4, 1403–1412.
\bibitem{8}
Corda C. Int. J. Mod. Phys. D 2009, 18, 2275–2282.
\bibitem{9}
M. Koussour; S. K. J. Pacif; M. Bennai and P.K. Sahoo. Fortschr. Phys. 2023, 2200172
\bibitem{10}
P. Shah and G. C. Samanta. Eur. Phys. J. C \textbf{79}, 414 (2019).
\bibitem{11}
Buchdahl H.A. Mon. Not. R. Astron. Soc. 1970, 150, 1–8.
\bibitem{12}
S. Capozziello, et al., Phys. Rev. D 2007, 76, 104019. https://doi.org/10.1103/PhysRevD.76.104019
\bibitem{13}
Harko T.; Lobo F.S.; Nojiri S.I.; Odintsov S.D. Phys. Rev. D 2011, 84, 024020.
\bibitem{14}
Sahoo P.K.; Moraes P.H.R.S.; Sahoo P.K. Eur. Phys. J. C 2018, 78, 46.
\bibitem{15}
Harko T.; Lobo F.S.N . Eur. Phys. J. C 2010, 70, 373–379.
\bibitem{16}
Jaybhaye L.V.; Solanki R.; Mandal S.; Sahoo P.K. Phys. Lett. B 2022, 831, 137148.
\bibitem{17}
Elizalde E.; Myrzakulov R.; Obukhov V.V.; Sáez-Gómez D. Class. Quantum Grav. 2010, 27, 095007.
\bibitem{18}
Bamba K.; Odintsov S.D.; Sebastiani L.; Zerbini S. Eur. Phys. J. C 2010, 67, 295–310.
\bibitem{19}
M. Koussour; M. Bennai; Class. Quantum Gravity 2022, 39, 105001. https://doi.org/10.1088/1361-6382/ac61ad
\bibitem{20}
Ferraro R.; Fiorini F. Phys. Rev. D 2007, 75, 084031.
\bibitem{21}
Myrzakulov R. Eur. Phys. J. C 2012, 71, 1752.
\bibitem{22}
J. B. Jimenez, et al., Phys. Rev. D 2018, 98, 044048. https://doi.org/10.1103/PhysRevD.98.044048
\bibitem{23}
J. B. Jimenez, et al., Phys. Rev. D 2020, 101, 103507. https://doi.org/10.1103/PhysRevD.101.103507
\bibitem{24}
Y. Xu, et al., Eur. Phys. J. C 2019, 79, 8. https://doi.org/10.1140/epjc/s10052-019-7207-4
\bibitem{25}
S. Mandal, et al., Phys. Rev. D 2020, 102, 024057. https://doi.org/10.1103/PhysRevD.102.024057
\bibitem{26}
S. Mandal, et al., Phys. Rev. D 2020, 102, 124029. https://doi.org/10.1103/PhysRevD.102.124029
\bibitem{27}
T. Harko, et al., Phys. Rev. D 2018, 98, 084043. https://doi.org/10.1103/PhysRevD.98.084043
\bibitem{28}
N. Dimakis, et al., Class. Quantum Grav. 2021, 38, 225003. https://doi.org/10.1088/1361-6382/ac2b09
\bibitem{29}
M. Koussour, et al., J. High Energy Astrophys. 2022, 35, 43. https://doi.org/10.1016/j.jheap.2022.05.002
\bibitem{30}
M. Koussour, et al., Phys. Dark Universe 2022, 36, 101051. https://doi.org/10.1016/j.dark.2022.101051
\bibitem{31}
N. Roy, Dynamical Systems Analysis of Various Dark Energy Models. PhD Thesis (2015). arXiv: 1511.07978 [gr-qc]
\bibitem{32}
S.D. Odintsov; V.K. Oikonomou, Phys. Rev. D 96, 104049 (2017)
\bibitem{33}
S.D. Odintsov, V.K. Oikonomou, P.V. Tretyakov, Phys. Rev. D 96, 044022 (2017)
\bibitem{34}
M. Hohmann, L. Jarv, U. Ualikhanova, Phys. Rev. D 96, 043508 (2017)
\bibitem{35}
A.S. Bhatia, S. Sur, Int. J. Mod. Phys. D 26, 1750149 (2017)
\bibitem{36}
K. Bamba, D. Momeni, M. Al Ajmi, Eur. Phys. J. C 78, 771 (2018)
\bibitem{37}
S. Carneiro, H. Borges, Gen. Relativ. Gravit. 50, 1 (2018)
\bibitem{38}
S. Santos Da Costa et al., Class. Quant. Grav. 35, 075013 (2018)
\bibitem{39}
P. Shah, G.C. Samanta, S. Capozziello, Int. J. Mod. Phys. A 33, 1850116 (2018)
\bibitem{40}
J. Wainwright, G.F.R. Ellis, Dynamical Systems in Cosmology (Cambridge University Press, Cambridge, 1997)
\bibitem{41}
A.A. Coley, Dynamical Systems and Cosmology (Springer, Netherlands, 2003)
\bibitem{42}
Chatzarakis, N., and V. K. Oikonomou. Annals of Physics 419 (2020) 168216.
\bibitem{43}
Odintsov, Sergei D., and Vasilis K. Oikonomou. Physical Review D 98.2 (2018) 024013.
\bibitem{44}
Odintsov, S. D., and V. K. Oikonomou. Classical and Quantum Gravity 36.6 (2019): 065008.
\bibitem{45}
Sebastian Bahamonde, Christian G. Bohmer, Sante Carloni, Edmund J. Copeland, Wei Fang, and Nicola Tamanini, Phys. Rept. 775-777, 1–122 (2018), arXiv:1712.03107 [gr-qc].
\bibitem{46}
[69] Edmund J. Copeland, Andrew R Liddle, and David Wands, Phys. Rev. D 57, 4686–4690 (1998), arXiv:gr-qc/9711068.
\bibitem{47}
Yungui Gong, Anzhong Wang, and Yuan-Zhong Zhang, Phys. Lett. B 636, 286–292 (2006), arXiv:gr-qc/0603050.
\bibitem{48}
M. R. Setare and E. N. Saridakis, Phys. Rev. D 79, 043005 (2009), arXiv:0810.4775 [astro-ph].
\bibitem{49}
Tonatiuh Matos, Jose-Ruben Luevano, Israel Quiros, L. Arturo Urena-Lopez, and Jose Alberto Vazquez, Phys. Rev. D 80, 123521 (2009), arXiv:0906.0396 [astro-ph.CO].
\bibitem{50}
Edmund J. Copeland, Shuntaro Mizuno, and Maryam Shaeri, Phys. Rev. D 79, 103515 (2009), arXiv:0904.0877 [astro-ph.CO].
\bibitem{51}
Yoelsy Leyva, Dania Gonzalez, Tame Gonzalez, Tonatiuh Matos, and Israel Quiros, Phys. Rev. D 80, 044026 (2009), arXiv:0909.0281 [gr-qc].
\bibitem{52}
Genly Leon and Emmanuel N. Saridakis, Class. Quant. Grav. 28, 065008 (2011), arXiv:1007.3956 [gr-qc].
\bibitem{53}
L. Arturo Urena-Lopez, JCAP 03, 035 (2012), arXiv:1108.4712 [astro-ph.CO].
\bibitem{54}
Genly Leon, Joel Saavedra, and Emmanuel N. Saridakis, Class. Quant. Grav. 30, 135001 (2013), arXiv:1301.7419 [astro-ph.CO].
\bibitem{55}
Carlos R. Fadragas, Genly Leon, and Emmanuel N. Saridakis, Class. Quant. Grav. 31, 075018 (2014), arXiv:1308.1658 [gr-qc].
\bibitem{56}
Maria A. Skugoreva, Emmanuel N. Saridakis, and Alexey V. Toporensky, Phys. Rev. D 91, 044023 (2015), arXiv:1412.1502 [gr-qc].
\bibitem{57}
Jibitesh Dutta, Wompherdeiki Khyllep, and Nicola Tamanini, Phys. Rev. D 93, 063004 (2016), arXiv:1602.06113 [gr-qc].
\bibitem{58}
Jibitesh Dutta, Wompherdeiki Khyllep, and Nicola Tamanini, Phys. Rev. D 95, 023515 (2017), arXiv:1701.00744 [gr-qc].
\bibitem{59}
Hmar Zonunmawia, Wompherdeiki Khyllep, Jibitesh Dutta, and Laur Jarv,Phys. Rev. D 98, 083532 (2018), arXiv:1810.03816 [gr-qc].
\bibitem{60}
Wompherdeiki Khyllep and Jibitesh Dutta, Eur. Phys. J. C 81, 774 (2021), arXiv:2102.04744 [gr-qc].
\bibitem{61}
Spyros Basilakos, Genly Leon, G. Papagiannopoulos, and Emmanuel N. Saridakis, Phys. Rev. D 100, 043524 (2019), arXiv:1904.01563 [gr-qc].
\bibitem{62}
Artur Alho, Claes Uggla, and John Wainwright, JCAP 09, 045 (2019), arXiv:1904.02463 [gr-qc].
\bibitem{texbook}
Ricardo G. Landim, Eur. Phys. J. C 79, 889 (2019), arXiv:1908.03657 [gr-qc].
\bibitem{63}
Wompherdeiki Khyllep, Jibitesh Dutta, Spyros Basilakos, and Emmanuel N. Saridakis, Phys. Rev. D 105, 043511 (2022), arXiv:2111.01268 [gr-qc].
\bibitem{64}
Jose Beltran Jimenez, Lavinia Heisenberg, and Tomi Koivisto, Phys. Rev. D 98, 044048 (2018), arXiv:1710.03116 [gr-qc].
\bibitem{65}
Jose Beltr an Jimenez, Lavinia Heisenberg, Tomi Sebastian Koivisto, and Simon Pekar,Phys. Rev. D 101, 103507 (2020), arXiv:1906.10027 [gr-qc].
\bibitem{66}
Wompherdeiki Khyllep, Jibitesh Dutta,  Emmanuel N. Saridakis and Kuralay Yesmakhanova, (2022) arXiv:2207.02610v1 [gr-qc]
\bibitem{67}
P. Shah and G. C. Samanta, Int. J. Mod. Phys. A \textbf{35}, 20501249 (2020).
\bibitem{68}
J. Beltran Jimenez. et. al., J. Cosmol. Astropart. Phys. 08 (2018) 039.
\bibitem{69}
Gaurav N. Gadbail, Simran Arora, P.k. Sahoo, Physics Letters B 838 (2023) 137710.
\bibitem{70}
Fotios K. Anagnostopoulos, Viktor Gakis, emmanuel N. Saridakis, and Spyros Basilakos, arXiv:2205.11445[gr-qc].
\bibitem{71}
Fotios K. Anagnostopoulos, Viktor Gakis, Emmanuel N. Saridakis, and Spyros Basilakos, (2022), arXiv:2205.11445 [gr-qc].
\bibitem{72}
Ruth Lazkoz, Francisco S. N. Lobo, Marıa OrtizBanos, and Vincenzo Salzano, Phys. Rev. D 100, 104027 (2019), arXiv:1907.13219 [gr-qc].
\bibitem{73}
Fotios K. Anagnostopoulos, Spyros Basilakos, and Emmanuel N. Saridakis, Phys. Lett. B 822, 136634 (2021), arXiv:2104.15123 [gr-qc].
\bibitem{74}
J. Wainright and G. F. R. Ellis, Dynamical Systems in Cosmology (Cambridge University Press, Cambridge, 1997).
\bibitem{75}
D. K. Arrowsmith and C. A. Place, An Introduction to Dynamical Systems (Cambridge University Press, Cambridge, 1990).
\bibitem{76}
M. W. Hirsch, R. L. Devaney and S. Smale, Differential Equations, Dynamical Systems and Introduction to Chaos (Academic Press, Elsevier, Amsterdam, 1974).
\bibitem{77}
S. Lefschetz, Differential Equations: Geometric Theory (Interscience, New York, 1957).
\bibitem{78}
S. Lynch, Dynamical Systems with Applications using Mathematica (Springer, Berlin, 2007).
\bibitem{79}
L. Perko, Differential Equations and Dynamical Systems (Springer, Berlin, 2001).
\bibitem{80}
S. Wiggins, Introduction to Applied Nonlinear Dynamical Systems and Chaos (Springer, 2003).
\bibitem{81}
J. Wainwright and G. F. R. Ellis, Dynamical Systems in Cosmology (Cambridge University Press, 2005).
\bibitem{82}
N. Tamanini, Phys. Rev. D \textbf{89}, 083521 (2014).
\bibitem{83}
Raja Solanki, Avik De, and P. K. Sahoo, arXiv:2203.03370 [gr-qc].
\bibitem{84}
S. A. Narawade, Laxmipriya Pati, B. Mishra, and S. K. Tripathy, arXiv:2203.14121 [gr-qc].
\bibitem{85}
Zinnat Hassan, Sayantan Ghosh, P.K Sahoo, Kazuharu Bamba, Eur. Phys. J.C (2022)82:1116.
\bibitem{86}
Bryan Sagredo, Savvas Nesseris, and Domenico Sapone, arXiv:1806.10822 [astro-ph.CO].
\bibitem{87}
Yong-Seon Song, Wayne Hu, and Ignacy Sawicki, arXiv:astro-ph/0610532.
\bibitem{88}
Gaurav N. Gadbail, Sanjay Mandal and Pradyumn Kumar Sahoo, Physics 2022,4,1403-1412.
\bibitem{89}
M. Koussour, S.K.j.Pacif, M. Bennai and P.K. Sahoo, Fortschr. Phys. 2023, 2200172.
\bibitem{90}
Elcio Abdalla et al., JHEAp 34, 49–211 (2022), arXiv:2203.06142 [astro-ph.CO].


\end{thebibliography}
\end{document}